**Vortex-induced vibration of a square cylinder in wind tunnel**

Xavier Amandolèse [a,b] , Pascal Hémon [c]*

[a] Département de Mécanique – Ecole Polytechnique, F-91128 Palaiseau Cedex, France
[b] Chaire d'Aérodynamique – CNAM, F-78210 Saint-Cyr l'Ecole, France
[c] LadHyX, CNRS – Ecole Polytechnique, F-91128 Palaiseau Cedex, France

* Corresponding author. Email pascal.hemon@ladhyx.polytechnique.fr

**Abstract**

An experimental study of the vortex-induced-vibration of a flexibly mounted rigid square cylinder in a uniform airflow is presented. For this high mass ratio configuration, transverse oscillations are measured in detail for reduced velocities ranging from 5 to 20. In the lock-in region and starting from rest, the cylinder motion exhibits a transient regime followed by a limit cycle oscillation regime. New experimental data are provided in term of amplitudes and frequencies of the limit cycle oscillations. The transient behaviour is also characterized by measuring the growth rate of the oscillations amplitude.

**1. Introduction**

Vortex shedding excitation of cylinder is probably one of the most studied problems in flow induced vibrations. This mechanism, referred to as Vortex-Induced Vibration (VIV), occurs on flexible cylinders when the vortices which develop in the wake can couple with the dynamics of the cylinder. It can be seen roughly as a resonance mechanism appearing when the frequency of the vortex shedding, controlled by the fluid flow, is close to the natural frequency of the cylinder. However the physics of the interactions between the flow and the cylinder transverse motion is not simply linear. The mathematical modelling of this problem in order to predict the amplitude of the cylinder motion has become a widely studied problem in engineering.

An extensive review of the vortex-induced vibrations of bluff bodies was published recently by Williamson and Govardhan [1] with a major focus on circular cylinder in water flow. Earlier, Bearman's contribution [2] and [3] were dealing with other bluff body shape such as the square section but mainly in the context of fixed or forced vibrations experiments. In all these publications and references therein, it appears that few experiments concern free vibrations of cylinder in airflow, mainly Feng [4] and Brika and Laneville [5] for a circular cylinder. Few data for a square section free to vibrate in airflow can be found in [6]. None of these publications presents data concerning transient regime.

The vortex shedding oscillations of high mass ratio structure (the mass ratio compares the mass of the cylinder to the surrounding fluid mass), is therefore not well documented, particularly for the square section, although its interest in civil engineering is obvious.

The objective of this paper is then to present new experimental results obtained in wind tunnel for an elastically mounted rigid square cylinder. Such measured data can serve for validation of predictive model. The behaviour of the vortex-induced oscillation is studied with the cylinder starting from rest, thus there is no hysteresis effect in the present study.

When the cylinder experiences vortex shedding oscillations, its response $z(t)$ exhibits a transient regime where the oscillations amplitude increases exponentially and a limit cycle oscillations regime (LCO) where the amplitude remains almost constant (see on Fig. 1). The objective of this paper is to present experimental values of the main characteristics of the vortex shedding excitation, *i.e.* the amplitude $\hat{z}$ of the LCO and its frequency $f$, and the growth rate of the oscillations amplitude in the transient regime, as a function of the reduced velocity.

## 2. Experimental techniques

### 2.1. Wind tunnel and flexible cylinder

Experiments are performed in a small vertical Eiffel-type wind tunnel with a closed circular test section of diameter 200 mm. A rigid square cylinder of spanwise length $L$ = 150 mm and breadth (cross section dimension) $D$=20 mm is elastically mounted (see Fig. 2) using four linear springs mounted outside the test section. Specific chord wiring is also used in order to restrain the cylinder to move transverse to the flow (see Fig. 2). This specific arrangement is suitably fitted in order to produce very low structural damping.

Mean-velocity and turbulence intensity distributions of the oncoming airflow have been measured. Over the velocity range of the experiments (ranging from 1.5 to 6 m/s) the wall-region of strong mean-velocity gradient is less than 15 mm. According to the length L of the cylinder (150 mm) compared to the diameter of the test section (200 mm) the cylinder is submitted to the flow in this core region where the non-uniformity of the mean-velocity is less than 5% and the turbulence intensity is less than 1 % over the velocity range of the experiments.

No endplates have been used in the experiments. Due to the aspect ratio of the cylinder (L/D=7.5) flow around the end of the cylinder could then have a significant effect on the vortex dynamics, the correlation of the induced fluid forces on the body and thus the vibrations. Meanwhile the proximity of both the ends of the cylinder with the test section wall could reduce the effect of end condition. Indeed, as reported by Morse et al [7] for a circular cylinder, vortex-induced vibration for attached and unattached endplates are nearly the same.

### 2.2. Measurement system

Downstream the test section, a nozzle is mounted in order to measure accurately the reference velocity of the wind tunnel, by using two sets of four static pressure taps, one in the test section and the second in the lowest section part of the nozzle. The mean flow velocity in the test section is deduced with Bernoulli's law between the two sections. Correction due to air temperature variation is performed using a thermocouple. This technique allows to measure the very low velocities needed in these experiments with an accuracy better than 1 %.

The transverse displacement $z(t)$ of the cylinder is measured by a laser displacement sensor. The measurement resolution is 40 μm and the accuracy is better than 1% over the full-scale range (± 10mm). The output signals are digitized with a 24 bits resolution acquisition system provided by Muller-BBM. The sampling resolution is 1024Hz and the duration of the acquisition is typically 60 seconds. For the frequency measurements of the LCO, this duration has been increased up to 300 seconds in order to obtain a better frequency resolution.

Preliminary tests have been performed in order to measure the Strouhal number of the cylinder at rest. Spectral analysis of the unsteady wake measurement was performed by a single component hot wire anemometer placed at a distance $D$ (20 mm) downstream the cylinder, at mid span, and slightly decentred in the transverse direction. The Strouhal number was found to be 0.127 over the velocity range of the vortex shedding oscillation regime, which is in accordance with Norberg's data for low Reynolds number [8].

### 2.3. Identification of structural parameters

In two dimensional configuration the equation of motion of the rigid cylinder reads

(1) $$m\ddot{z} + c\dot{z} + kz = F,$$

where $m$ is the mass of the cylinder, $c$ is a viscous type damping coefficient associated with the springs, the chord wiring and their mounting, $k$ is the stiffness of the set-up and $F$ is the time-dependent aerodynamic force resulting from the fluid force, mainly that due to vortex shedding. Equation (1) can also be written as following

(2) $$\ddot{z} + 2\eta(2\pi f_0)\dot{z} + (2\pi f_0)^2 z = \frac{F}{m},$$

where the undamped natural frequency of the system is

(3) $$f_0 = \frac{1}{2\pi}\sqrt{\frac{k}{m}}$$

and $\eta = \dfrac{c}{2\sqrt{km}}$ is the damping ratio.

The structural parameters of the system are determined experimentally without wind. The stiffness $k$ is measured by static calibration using reference masses. The natural frequency $f_0$ and the structural damping $c$ are measured by spectral analysis of free vibration responses to transient deflections. According to the very low damping of the system, the mass $m$ can be deduced from the stiffness $k$ and the natural frequency $f_0$ using the relation (3).

Results are reported in Table I along with the geometric dimensions of the cylinder, the physical parameters of the airflow and the velocity range of the experiments.

Pertinent non dimensional parameters are reported in Table II. One can notice that the system has a high mass ratio $m^* \approx 905$ associated with a very low damping ratio $\eta \approx 0.0828$ %. The very low damping leads to a relatively small Scruton number $Sc$ close to 1.5 which is the key parameter in the observation of vortex shedding vibrations.

## 3. Limit cycle oscillations

### 3.1. Amplitudes

High mass ratio cylinder exhibits vortex-induced oscillations for which the maximum amplitude of the LCO is expected for reduced velocity close to $1/St$. For the square cylinder it should then occurs at $Ur \approx 8$.

Starting from rest, LCO of the square cylinder are measured for reduced velocity ranging from $Ur = 2$ up to 26. The reduced RMS amplitude of the limit cycle oscillation $Z^* = \hat{z}/D$ is presented in Figure 3 as a function of the reduced velocity. For reduced velocity below 6 no significant oscillation occurs. For $Ur$ ranging from 6 up to 13, a typical VIV amplitude response can be observed. At higher reduced velocity galloping oscillations appear which are not studied here.

In the VIV regime the amplitude data shown in Figure 3 are very similar to those carried out by Feng [4] for a circular cylinder in airflow. Indeed those results clearly show two amplitude branches, which, according to Khalak and Williamson [9], could be named the "initial" branch and the "lower" branch. Due to the specific arrangement of the initial and lower branches it seems that, as for the experiments conducted by Feng [4], a higher amplitude will be achieved by increasing the reduced velocity over a certain range on the initial branch than in decreasing back the reduced velocity over the same range on the lower branch.

One can also notice that the initial and the lower branches cross each over for a reduced frequency close to 8 ($\approx 1/St$), which is the expected pure resonant point. Meanwhile the maximum oscillation amplitude occurs on the initial branch for a reduced velocity close to 9.

The present experiments have been performed for a cylinder starting from rest and hysteretic transition between branches has not been observed. Meanwhile long time analysis of the LCO regime clearly showed intermittent switching between the initial and the lower branches for reduced velocity ranging from 8 to 9. In the data provided in Figure 3, the acquisition time was chosen so that only one kind of regime is recorded for a given test.

### 3.2. Frequencies

The reduced frequency of the LCO ($F^* = f/f_0$) is presented in figures 4 and 5 as a function of the reduced velocity. On those figures the reduced frequency associated with the vortex shedding frequency $f_w$ based on the Strouhal number ($St\,Ur = f_w/f_0$) is also reported.

Zooming the evolution of the reduced frequency as a function of the reduced velocity around lock-in (see on Fig. 5) one can observe the frequency evolution of the oscillations. At the beginning of the lock-in for $Ur < 8$, the oscillations frequency first decreases significantly in the direction of the vortex shedding frequency. It then increases between $Ur = 8$ to 11 to reach a value slightly upper the natural frequency of the cylinder

This behaviour is very close to the one underlying by de Langre [10] using linear VIV dynamic systems. According to this author, this is caused by classical couple mode-flutter mechanism between the cylinder dynamics and the wake dynamics.

Reduced frequencies reported on figure 5 are the result of long time analysis in the LCO regime. In that context frequencies for $Ur < 8$ can clearly be associated with the initial branch. As for frequencies for $Ur > 9$ which can be associated with the lower branch. For reduced velocity ranging from 8 to 9 it is not so clear due to the intermittent switching between the initial and the lower branches.

## 4. Transient behaviour

In the transient regime the square cylinder oscillation amplitude increases quasi exponentially (see Fig. 1). Therefore its envelope can be fitted by an expression as following: $A \times \exp(\lambda(t - t_0))$. In that context the growth rate of oscillations amplitude in the transient regime is defined as $\delta \equiv \dfrac{\lambda}{2\pi f_0}$ so as to be expressed as a ratio of the critical damping $c_c$.

The growth rate $\delta$ has been identified in the VIV regime for reduced velocity ranging from 7 up to 14. It must be noted that the growth rate values that are reported in Figure 6 have not been corrected by the damping ratio of the cylinder motion in still fluid. To do so and to express a growth rate due to pure aerodynamics effect one has to subtract the damping ratio value $\eta = 0.0828$ % to the growth rate data presented in Figure 6

Results show a sharp increase at the beginning of the lock-in, with a maximum slightly above 0.2% for a reduced velocity corresponding to the matching of the oscillations frequency with the vortex shedding frequency ($Ur \approx 1/St \approx 8$). Beyond, the growth rate then decreases in a slightly smoother way. As for frequencies, this growth-rate behaviour can be highlighted using classical couple mode-flutter analysis between the cylinder dynamics and the wake dynamics. This has been theoretically reported by de Langre [10] using linearised VIV dynamic modelling. Meanwhile a hysteretic behaviour seems to occur for reduced velocity ranging from 8 to 9, which would need to be highlighted in a further step.

## 5. Conclusion

An experimental study on the vortex-induced transverse oscillation of a flexibly mounted rigid square cylinder in a uniform airflow has been presented. For this high mass ratio configuration with very low "structural" damping new experimental data have been provided in term of amplitude and frequency of limit cycle oscillations. Growth rate of the oscillation amplitude in the transient regime have also been measured as a function of the reduced velocity.

An important feature is the similarity of behaviour of the square section cylinder with that of the circular section cylinder. Therefore some questions still remain for this simple configuration of elastically mounted square cylinder. Especially the hysteretic transition between the initial and the lower branch has to be highlighted. So does the possible change in the wake pattern associated with the jump from the initial to the lower branch (as reported by Brika and Laneville [5] for a circular cylinder).

**References**

[1] C.H.K. Williamson and R. Govardhan, Vortex Induced Vibrations, Annu. Rev. Fluid Mech. 36 (2004) 413-455.

[2] P.W. Bearman, E.D. Obasaju, An experimental study of pressure fluctuations on fixed and oscillating square-section cylinders, J. of Fluid Mech. 119 (1982) 297-321.

[3] P.W. Bearman, Vortex shedding from oscillating bluff bodies, Annu. Rev. Fluid Mech. 16 (1984) 195-222.

[4] C.C. Feng, The measurement of vortex induced effects in flow past stationary and oscillating circular and d-section cylinders, Master's Thesis, Department of Mechanical Engineering, The University of British Columbia, Canada, 1968.

[5] D. Brika and A. Laneville, Vortex-induced vibrations of a long flexible cylinder, J. of Fluid Mech., 250 (1993) 481-508.

[6] L. Cheng, Y. Zhou, M.M. Zhang, Perturbed interaction between vortex shedding and induced vibration, J. of Fluids and Structures, 17 (2003) 887-901.

[7] T.L. Morse, R.N. Govardhan, C.H.K. Williamson, The effect of end conditions on the vortex-induced vibration of cylinders, J. of Fluids and Structures, 24 (2008) 1227-1239.

[8] C. Norberg, Flow around rectangular cylinders: Pressure forces and wake frequencies, J. of Wind Engineering and Industrial Aerodynamics, 49 (1993) 187-196.

[9] A. Khalak, C.H.K. Williamson, Dynamics of a hydroelastic cylinder with very low mass and damping, J. of Fluids and Structures, 10 (1996) 455-472.

[10] E. de Langre, Frequency lock-in is caused by coupled-mode flutter, J. of Fluids and Structures, 22 (2006) 783-791.

**Figures captions**

*Figure 1. Time evolution of the cylinder motion amplitude at U=2.5155 m/s*

*Figure 2. Sketch showing the principles of the experimental setup*

*Figure 3. Reduced rms amplitude of the limit cycle oscillations versus reduced velocity*

*Figure 4. Reduced frequency of the limit cycle oscillations versus reduced velocity;
▫ measurements; ─ reduced vortex shedding frequency St Ur*

*Figure 5. Reduced frequency of the limit cycle oscillations versus reduced velocity; same as Figure 4 zoomed around the lock-in region*

*Figure 6. Growth rate of the oscillations (percentage of the critical damping) versus reduced velocity*

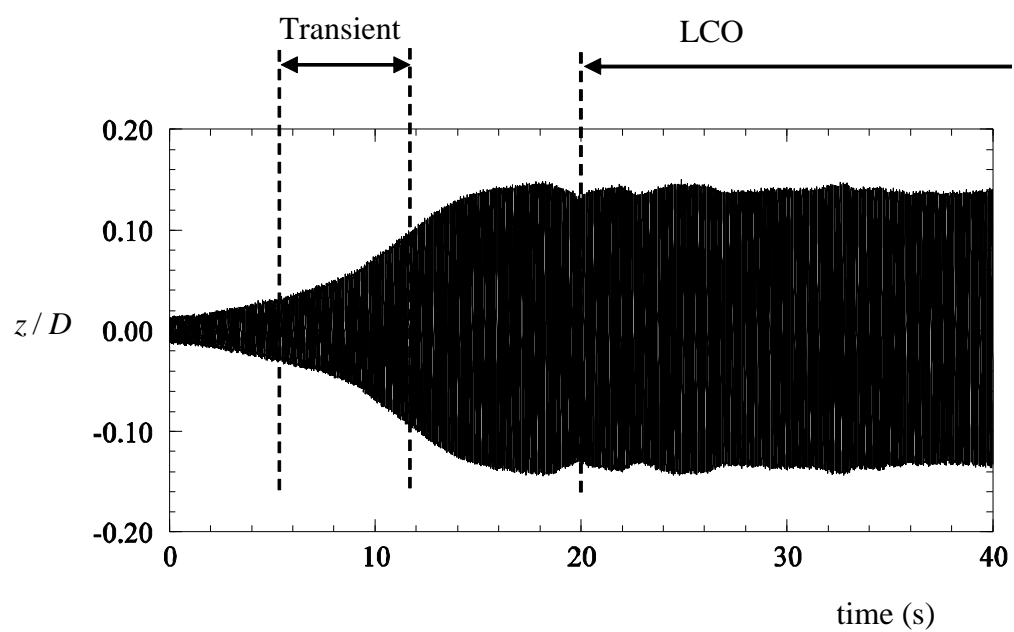

Figure 1. Time evolution of the cylinder motion amplitude at U=2.5155 m/s

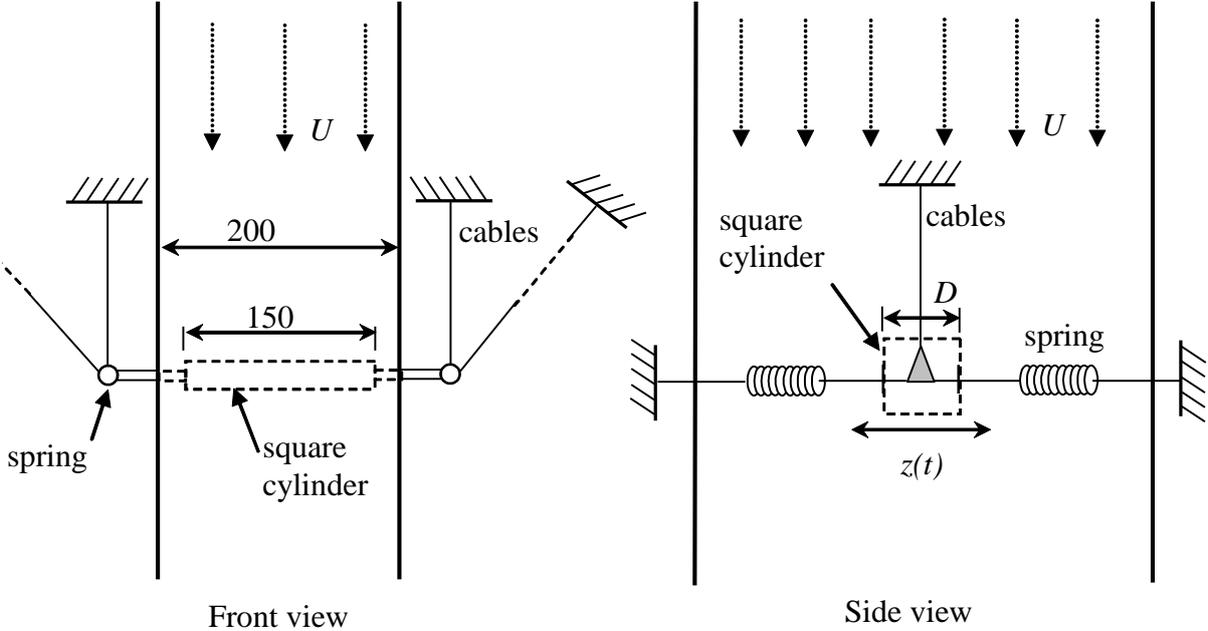

*Figure 2. Sketch showing the principles of the experimental setup*

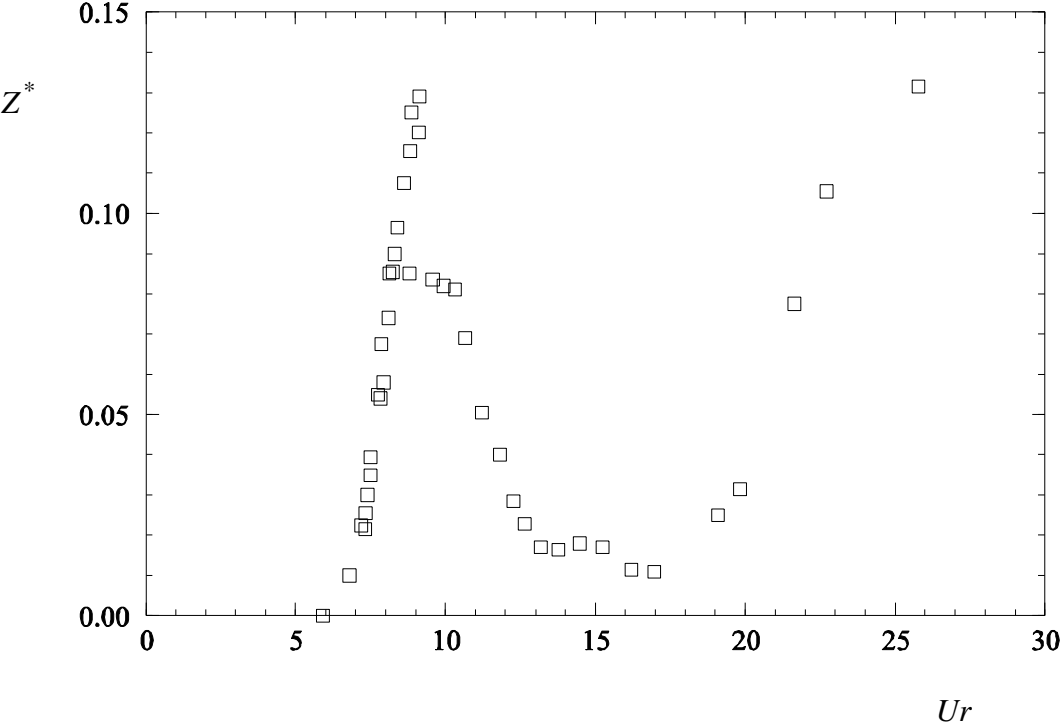

*Figure 3. Reduced rms amplitude of the limit cycle oscillations versus reduced velocity*

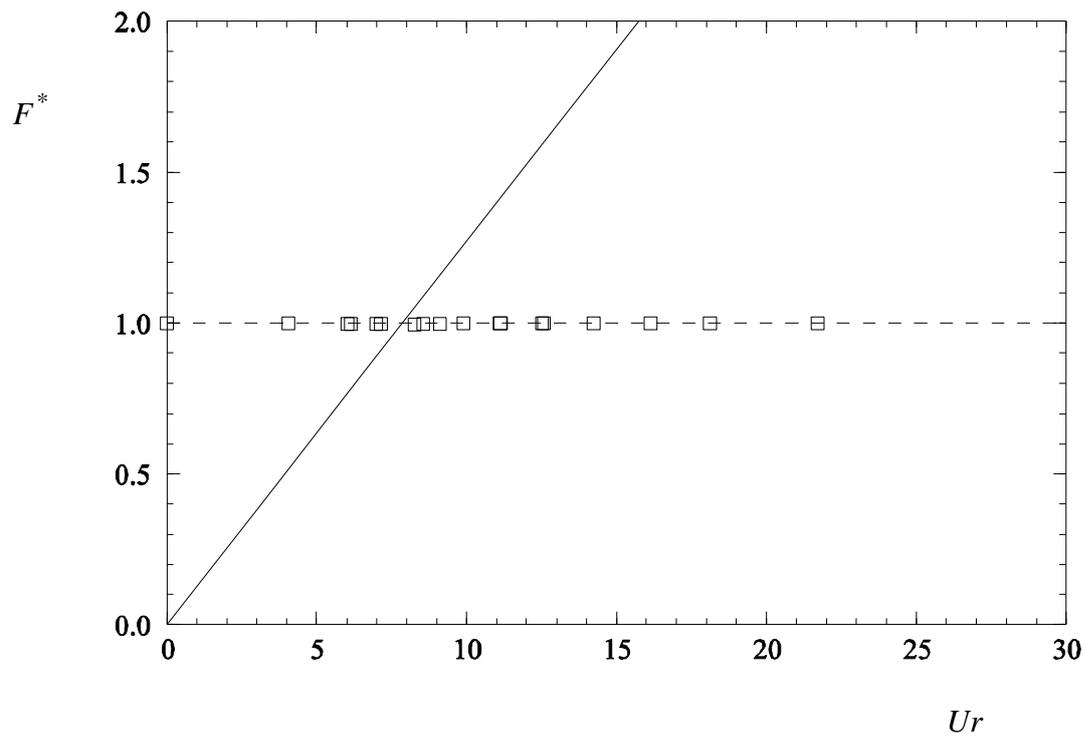

*Figure 4. Reduced frequency of the limit cycle oscillations versus reduced velocity;*
*☐ measurements; — reduced vortex shedding frequency St Ur*

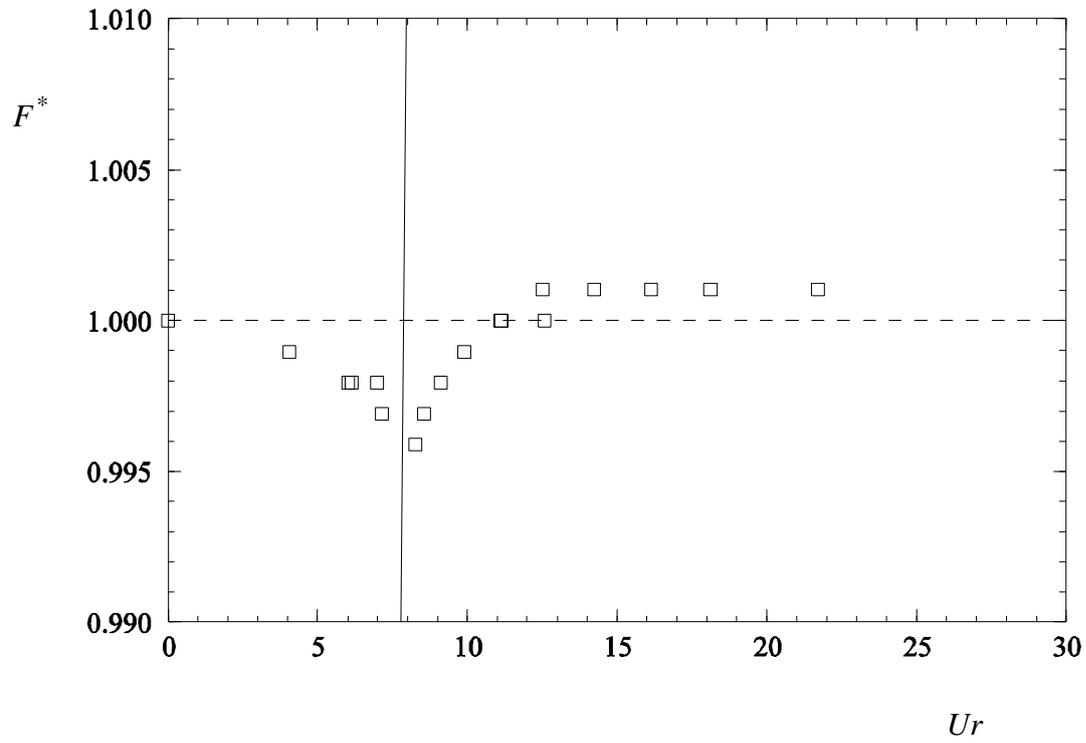

*Figure 5. Reduced frequency of the limit cycle oscillations versus reduced velocity; same as Figure 4 zoomed around the lock-in region*

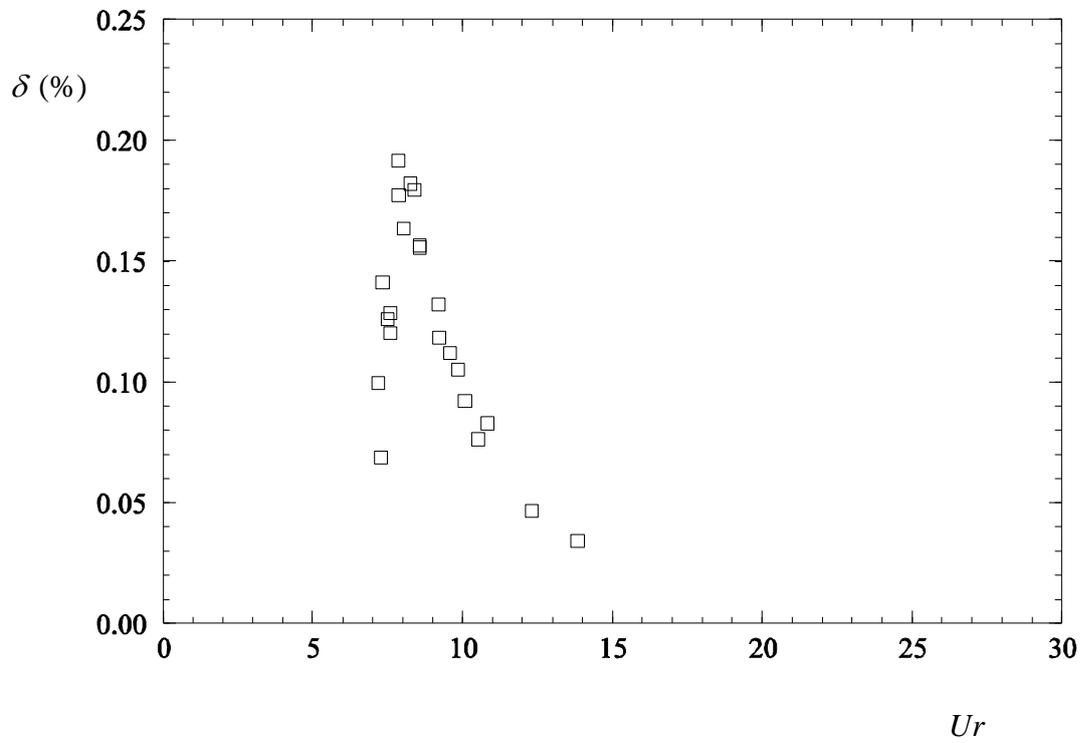

*Figure 6. Growth rate of the oscillations (percentage of the critical damping) versus reduced velocity*

Table I. Physical parameters of the experiments

| | | | |
|---|---|---|---|
| Diameter of the cylinder | $D$ | 20 | mm |
| Length of the cylinder | $L$ | 150 | mm |
| Stiffness of the setup | $k$ | 597.6 ± 35 | N/m |
| Mass of the cylinder | $m$ | 0.0654 ± 0.004 | kg |
| Critical damping | $c_c$ | 12.5 ± 0.75 | N.s/m |
| Structural damping | $c$ | 0.0104 ± 0.0008 | N.s/m |
| Natural frequency | $f_0$ | 15.21875 ± 0.01563 | Hz |
| Wind tunnel velocity | $U$ | 1.5 – 6.0 | m/s |
| Air density | $\rho$ | 1.205 | kg/m$^3$ |
| Kinematic viscosity | $\nu$ | 15 10$^{-6}$ | m$^2$/s |

Table II. Non dimensional parameters

| | | | |
|---|---|---|---|
| Reynolds number | $Re$ | $\dfrac{U D}{\nu}$ | 2000 - 8000 |
| Mass ratio | $m^*$ | $\dfrac{m}{\rho D^2 L}$ | 905 |
| Damping ratio | $\eta$ | $\dfrac{c}{c_c}$ | 0.000828 ± 0.000014 |
| Scruton number | $Sc$ | $2 \eta m^*$ | 1.498 |
| Strouhal number | $St$ | $\dfrac{f_w D}{U}$ | 0.127 |
| Skop-Griffin parameter | $S_G$ | $4\pi^2 St^2 Sc$ | 0.954 |
| Reduced velocity | $Ur$ | $\dfrac{U}{f_0 D}$ | 5 – 20 |